\documentclass[useAMS,usenatbib,onecolumn]{mn2e}
\usepackage{graphicx}

\newcommand{\eqref}[1]{(\ref{#1})}

\newcommand{\3}{$_{3}$}

\newcommand{\cm}{cm$^{-1}$}

\newcommand{\p}{^\prime}
\newcommand{\pp}{^{\prime\prime}}

\newcommand{\Dh}{${\mathcal D}_{3{\rm h}}$}

\title[Hot line list for Ammonia]{ A variationally computed line list for hot NH\3}

\author[S. N. Yurchenko, R. J. Barber, J. Tennyson]{S. N. Yurchenko$^1$, R. J. Barber$^2$, J. Tennyson$^2$ \\
$^1$Technische Universit\"at Dresden, Institut f\"ur
Physikalische Chemie und Elektrochemie, D-01062 Dresden, Germany.\\
$^2$Department of Physics and Astronomy, University
College London, Gower Street, WC1E 6BT London, UK}

\date{Accepted XXXX. Received XXXX; in original form XXXX}

\pagerange{\pageref{firstpage}--\pageref{lastpage}} \pubyear{2010}

\begin{document}

\label{firstpage}

\maketitle

\begin{abstract}
We present `BYTe', a comprehensive `hot' line list for the
ro-vibrational transitions of ammonia, $^{14}$NH$_3$, in its ground
electronic state. This line list has been computed variationally using
the program suite TROVE, a new spectroscopically-determined potential
energy surface and an \textit{ab initio} dipole moment surface.  BYTe,
is designed to be used at all temperatures up to 1\,500~K. It
comprises 1\,137\,650\,964 transitions in the frequency range from 0
to 12\,000~cm$^{-1}$, constructed from 1\,366\,519 energy levels below
18\,000~cm$^{-1}$ having $J$ values $\leq 36$. Comparisons with
laboratory data confirm the accuracy of the line list which is
suitable for modelling a variety of astrophysical problems including
the atmospheres of extrasolar planets and brown dwarfs.

\end{abstract}


\maketitle

\section{Introduction}
\label{sec:intro}

Ammonia is the main nitrogen-containing molecule observable in a variety
of astrophysical environments.  For example, whilst the spectra of M
and L-type brown dwarfs are dominated by H$_2$O, and CH$_4$ becomes
increasingly important at lower temperatures, NH$_3$ is a significant
source of opacity in the atmospheres of late T-type dwarfs,
particularly in the 10.5 $\mu$m region \citep{Burrows_etal_97,
Sharp_Burrows_07}. Modelling suggests that absorption by ammonia will
be even more important in the yet-to-be discovered Y-dwarfs
\citep{Burrows_etal_2003} and it is possible that NH$_3$ absorption
bands may be the principal distinguishing feature of this new class of
ultra-cool dwarf.

The molecule is known to be present in the atmosphere of solar system
gas giant planets \citep{Lara_etal_98}. For example, emission spectra
of hot ammonia were observed in the atmosphere of Jupiter after the
impact of Comet Shoemaker-Levy 9 \citep{jt154} and in the
aftermath of the July 2009 impact on Jupiter \citep{Wesley1,Wesley2}.

NH\3\ is present in cometary comae, typically in number densities
$\sim 0.5\%$ of that of gaseous H$_2$O \citep{kw02,
Bonev_etal_09}. Although the cometary environment is cold, solar
pumping \citep{jt330,km10} and other excitation mechanisms \citep{jt452}
are able to excite high lying ro-vibrational molecular states that in
conditions of thermal equilibrium would only be significantly
populated at temperatures in the region of 2\,000 K or higher. In the
densities that exist in the coma, these states are able to decay
radiatively before being collisionally de-excited, and spectral lines
originating from these highly excited states give important clues to
the physical processes at work in cometary comae. The Einstein A
coefficient listed in BYTe will enable the measured intensities of
cometary NH$_3$ spectral lines to be used in modelling the conditions
in the coma.

Ammonia masers are also observed. These also involve transitions
between high lying states \citep{mim86}. Modelling maser
action requires significant spectroscopic data, and this too can be provided
by our line list, which we call BYTe.

An exciting new area of astronomy is the characterisation of extrasolar
planets. Water and methane have already been shown to be
present in exoplanet atmospheres (for example in HD~189733b
\citep{jt400,Swain_etal_2008}), but to-date NH$_3$ has not. The reactions by
which N$_2$ and H$_2$ are converted into ammonia in the atmospheres of
expoplanets and brown dwarfs are complex and outside the scope of this
paper. However, the equilibrium between N$_2$ and NH$_3$ favours
NH$_3$ at lower temperatures as (to a lesser extent) do higher
pressures. These temperature and pressure dependencies suggest that
the outer atmospheres of extrasolar giant planets  at large
orbital distances will contain significant quantities of
ammonia \citep{Sudarsky_etal_03}, and it is likely that NH$_3$ will
soon be confirmed to be present in these objects.

Before BYTe, there was no NH$_3$ line list that was sufficiently
complete and accurate for use in modelling the atmospheres of
exoplanets, brown dwarfs or other astronomical objects at elevated
temperatures. Many astronomers have used the experimental NH\3\ line
list of \citet{NH3-Irwin}, which covers the 4\,000-11\,000~\cm\
spectral region, even though this line list is only designed to be
used at temperatures below 300~K.  The HITRAN database~\citep{HITRAN}
has been also extensively used. This contains approximately 30~000
lines of $^{14}$NH$_3$, but it too is only suitable for use at ambient
temperature, or below, and even then is lacking data in significant
regions of the spectrum; we recently demonstrated this using our
computed, comprehensive line list for cold NH\3\
\citep{NH3-T300K-paper}. This line list comprises 3.25 million
transitions between 184\,400 levels. It has an upper energy cut-off of
12\,000 \cm\ and a maximum rotational quantum number $J$ =
20. However, despite having many more lines than any other NH$_3$
list, like those other lists, it was designed for use at temperatures
up to 300 K, which renders it unsuitable for most astronomical
applications. This problem is overcome by BYTe.

\section{The BYTe line list}

BYTe is a catalogue of transitions represented by frequencies,
Einstein coefficients, energy levels, and quantum numbers, which
together fully characterise the electric dipole transitions of the
ammonium molecule in the frequency range from 0 to
12\,000~cm$^{-1}$. The list was computed variationally using the
program suite TROVE~\citep{trove-paper}. It comprises 1\,137\,650\,964
transitions between 1\,366\,519 energy levels, all below
18\,000~cm$^{-1}$, and for all $J$s up to $J=36$. All transitions are
within the ground electronic state of the $^{14}$NH$_3$, since the
excited electronic states are above the dissociation limit of the
molecule. Less than 30\,000 NH\3\ lines are known experimentally: BYTe
contains about 40\,000 times as many. It is the most comprehensive and
accurate line list of NH$_3$ and is designed to be applicable at
temperatures up to 1\,500~K.

The ro-vibrational energies and wave functions were computed using a
new `spectroscopic' potential energy surface (PES) of
NH$_3$~\citep{PES-NH3-2010}. This PES was recently generated through a
least-squares fit to the $^{14}$NH$_3$ experimental energy levels
mostly taken from the HITRAN database~\citep{HITRAN}. In order to
evaluate the Einstein coefficients an \textit{ab initio} (ATVZ) dipole
moment surface of NH$_3$ from \citet{Yurchenko-NH3-DMS-2005} is
used. The reported `hot' NH\3\ line list is the result of more than
three years work by our team. The method of calculation was similar
to, but more computer-intensive than in our earlier $T$ = 300~K NH\3\
line list~\citep{NH3-T300K-paper}.

Preliminary versions of BYTe have already been used in several
astrophysical studies. In particular the data has been used in
analysing the atmosphere of the late T dwarf
UGPS~0722-05~\citep{Brown-dwarf} and the atmosphere of the transiting
hot Neptune GJ436b~\citep{Neptune}

\section{Computational details}
\label{s:comput-details}

The variational rotation-vibration program suite
TROVE~\citep{trove-paper} was employed for all nuclear motion
calculations required to produce the BYTe line list. This procedure
was the same as that used in our cold line
list~\citep{NH3-T300K-paper}, and the reader is referred to this
earlier paper for a more detailed account of TROVE. The main
difference in generating BYTe was that a larger basis set was used
than in the case of the cold NH$_3$ paper.

TROVE normally requires a polyad number $P$ to control the basis
set. In the case of ammonia, we define:
\begin{equation}\label{e:polyad-2}
    P =  2 (n_1 + n_2 +n_3) + n_{4} + n_{5} + \frac{n_{6}}{2},
\end{equation}
where $n_i$ are the quantum numbers associated with the basis functions
$\phi_{n_i}$, see \citet{NH3-T300K-paper}. Thus, we include in the
basis set only those functions $\phi_n$ for which $P$ $\leq$ $P_{\rm max}$.
$P_{\rm max}$ was set at 14. This optimum value produced converged
eigenvalues with the minimum computational demands.

The Hamiltonian matrices were constructed from the ($J$=0)-contracted basis
set~\citep{NH3-T300K-paper}.  In order to improve the agreement of
our line frequencies with experiment, the empirical basis set
correction (EBSC) was utilized~\citep{NH3-T300K-paper}. Using the EBSC
approach, the vibrational energies in the ro-vibrational calculations
were replaced by the corresponding experimental band centers, where
these are available.

Apart from extending the basis set, we also enlarged (i) the energy
range from $E_{\rm max} = 12\,000$~\cm\ to  $E_{\rm max} =
18\,000$~\cm, (ii) the frequency range from $0$ -- $8\,000$~\cm\ to
$0$ -- $12\,000$~\cm, and (iii) the range of the rotational
excitations considered from $J_{\rm max}=20$ to $J_{\rm max}=41$,
respectively. These changes were necessary for the new line list to
be sufficiently complete and accurate to enable spectral calculations
at temperatures up to $T=1500$~K, which is far more demanding than at
$T=300$~K, which was the temperature criterion adopted in
\citet{NH3-T300K-paper}.

\begin{figure}
\begin{center}
\includegraphics[width=0.6\linewidth]{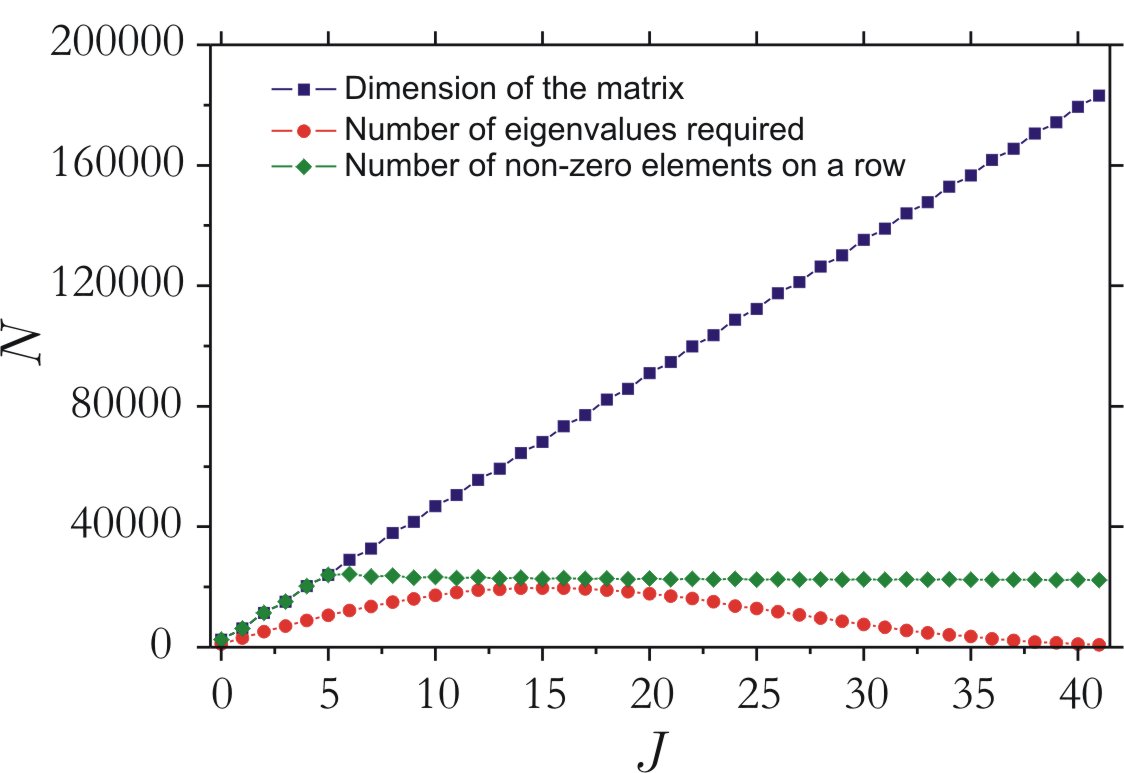}
\end{center}
\caption{\label{f:Nsizes:Nroots} Dimensions of the $E$-symmetry matrices (squares), the corresponding number of eigenvalues below 18\,000~\cm (circles) and
number of non-zero elements on each row (diamonds). }
\end{figure}

The extensions to these key parameters has a huge impact on the
computational costs. In the case of BYTe, the largest Hamiltonian
matrix to be diagonalized was for $J=36$ ($E$ symmetry block) with
dimension 162\,763. In contrast, the largest Hamiltonian in
\citet{NH3-T300K-paper} was  86\,000 at $J=20$. However, the  most elaborate calculations,
were in the $J$ = 20--27 range. These were characterised not only by large
matrix dimensions (80\,000 -- 110\,000), but also by a large number of
roots of the corresponding eigenvalue problems. This is illustrated on
Fig.~\ref{f:Nsizes:Nroots}, where we show the number of eigenvalues
below the 18\,000~\cm\ threshold and the dimensions of the
$E$-symmetry matrices for $J = 0 \ldots 41$. Using the computer facilities
available to us we could efficiently employ the LAPACK diagonalizer
dsyev for all $J$ below 23. For $J=24$ and higher we had to switch to
the Lanzcos-based iterative diagonalizer PARPACK~\citep{ARPACK} (a
parallelized version of ARPACK). In fact, the dsyev LAPACK routine
allowed us to compute and store all eigenvalues of a given Hamiltonian
matrix for $J\le 23$, beyond the $E_{\rm max} = 18\,000$~\cm\
threshold.  One can see from Fig.~\ref{f:Nsizes:Nroots} that for
$J > 30$ the number roots to be determined in the eigenvalue problem
represents only a small fraction of the matrix dimension (less than
10~\%). This is important for efficient utilization of the
iterative Lanzcos-type diagonalization approaches. The matrix
diagonalization is the most consuming part of line list calculation,
both in terms of time and memory.

Using these techniques, we were able to generate all
eigenvalues and eigenvectors for values of $J$ up to 41, subject to the energy
threshold of $18\,000$~\cm. However, only levels below $J=36$ were
taken into the line list because higher $J$ values did not have any
ro-vibrational states with energies below the threshold for the lower
energy states in BYTe of $E_{\rm low}$ = $8\,000$~\cm.

As previously mentioned, in order to compute the eigenvalues and
eigenvectors of NH\3\ we employed a new, improved, `spectroscopic'
PES, NH3-2010.  This PES was recently generated through fitting to the
available experimental data below $J=8$. It involved adjustments to
the analytical representation from \citet{Yurchenko2005-NH3:survey},
using the refined PES from this work as a starting point.  This
refinement was performed using TROVE, which was extended to allow such
fitting tasks. The details of the refinement procedure and the new PES
NH3-2010 will be given elsewhere~\citep{PES-NH3-2010}.

An accurate dipole moment surface (DMS) is a prerequisite for
producing accurate line intensities.  Tests have shown that these are
best taken directly from high quality {\it ab initio} calculations
\citep{jt156}.  For BYTe we employed the ATZfc DMS of NH\3\ from
\citet{Yurchenko-NH3-DMS-2005} in the improved DMS representation from
\citet{NH3-T300K-paper}, which behaves well when the molecule adopts a
planar geometry.  This DMS has been shown to provide intensities in
good agreement with experiment
\citep{Yurchenko-NH3-DMS-2005,NH3-T300K-paper}; it was also used in
producing the `cool' ammonia line list~\citep{NH3-T300K-paper}.

To speed up the computation of the transition moments used to generate
the Einstein A coefficients, we used the pre-screening procedure, see
\citet{NH3-T300K-paper}, in which only eigen-coefficients with
magnitude larger than $10^{-12}$ were selected. This reduced the size
of the vectors (by about 70-80\%) as well as the computation time. We
also imposed an absorption intensity threshold $10^{-12}$~cm$/$~mol
(1.7$\times 10^{-36}$~cm/molecule) at $T=1\,500$~K, for the intensities
to be included in BYTe. This threshold corresponds to about $10^{-16}$
of the maximum intensity at $T=1\,500$~K (18\,000~cm$/$mol). The major
bottle-neck in these calculations was associated with the need to read
repeatedly a huge number of eigenvectors stored on the disk. The large
size of these vectors prevented us from keeping all of them in the
virtual memory during the computational process.  We were able to
reduce the number of readings by optimizing the computational
logistics. In intensity simulations each transition can be
independently processed, which effectively naturally parallelizes
them, distributing them between computational nodes (see
\citet{NH3-T300K-paper}).

\begin{figure}
\begin{center}
\includegraphics[width=0.6\linewidth]{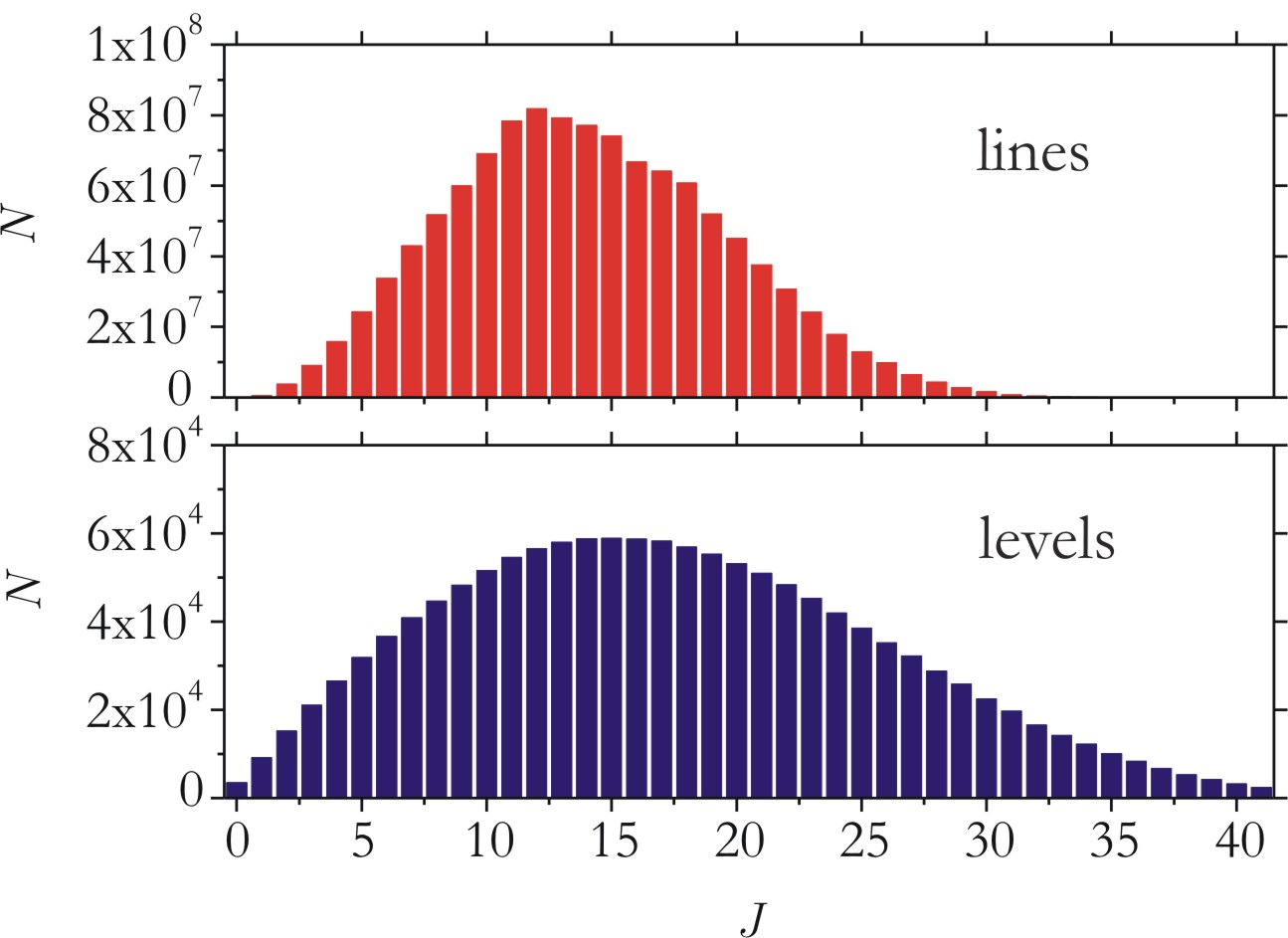}
\end{center}
\caption{\label{f:Nlevels:Nlines} Numbers of levels below 18\,000 \cm\ per each $J$ (lower display)
and corresponding transitions for each $J \to J+1$ (upper display).  }
\end{figure}

The most expensive part of the intensity calculations was for $J$ from
$10$ to $20$, which accounts for approximately 75~\% of all transitions to
be computed.  In Fig.~\ref{f:Nlevels:Nlines} we show how the
number of lines and energy levels depend on $J$. The number of levels
peaks at $J=15$ and the reduces gradually with higher $J$ and eventually
reaches zero due to our lower and upper energy limits of $8\,000$ and
$18\,000$~\cm\ respectively.

\section{Structure of the line list of NH\3}
\label{s:linelist}

The BYTe $^{14}$NH\3\ line list contains 1.138 billion lines, which
are the allowed transitions above a certain minimum intensity, between
1.367 million ro-vibrational levels. The structure of BYTe is similar
to that of the BT2 water line list \citep{BT2}. BYTe comprises two
files. One, the Energy file, holds the energies and ro-vibrational
quantum numbers of all NH$_3$ states up to $J\le 41$ that are less
than 18\,000~\cm\ above the zero point energy (our value is
7430.288276~\cm). Table~\ref{t:Energy-file} gives an extract from the
Energy file. Apart from the general quantum numbers associated with
the molecular group symmetry $\Gamma$~\citep{TheBook} (NH$_3$ belongs
to \Dh(M)) and total angular momentum $J$, we have used both `normal
mode' and `local mode' quantum numbers when labelling the energy
levels. Our labelling scheme is discussed below.

The structure of the Transition file is simpler. It contains three
columns: two give the reference numbers of the upper and lower states
as they appear in the Energy file and third contains the Einstein A
coefficient (s$^{-1}$) for the transition (see
Table~\ref{t:Transit-file}). The entries in the Transition files are
sorted according to the frequency, and we have split them into 120
small files in order to reduce the amount of data that needs
to be handled when examining a specific frequency region.

In the actual TROVE calculations we employed `local mode' basis
functions in the FBR representation as explained in detail in
\citet{NH3-T300K-paper}. This allowed us to label the energy levels
and hence to assign transitions based on the particular basis set
making the largest contribution within the appropriate
eigenfunction. Our local mode quantum numbers include $\Gamma_{\rm
rot}, K, \tau_{\rm rot}$, $\Gamma_{\rm vib}$, $n_1, n_2, n_3, n_4,
n_5, n_6$. Here $K$ is the projection of total angular momentum onto the
molecular symmetry axis; $n_1,\ n_2,\ n_3$ are stretching local mode
quantum numbers \citep{mr85} which correlate with the normal mode
notation as $n_1 + n_2 + n_3 = \nu_1 + \nu_3$; $n_4$ and $n_5$ are
deformational bending quanta; $n_6$ is the inversion quantum number
equivalent to $2 v_2 + \tau_{\rm inv}$, where $v_2$ is the normal mode
quantum number and $\tau_{\rm inv} = n_6 \ mod \ 2$ is the inversion
parity~\citep{MolPhysPaper}. Finally, $\Gamma_{\rm rot}$ and
$\Gamma_{\rm vib}$ are the rotational and vibrational symmetries in
\Dh(M).

Apart from the `local mode' assignment that is generated by TROVE, we
also provide the standard normal mode quantum numbers $v_1, v_2, v_3^{l_3},
v_4^{l_4}$, according with the Herzberg convention \citep{Herzberg_45}. $v_1$
and $v_2$ are the symmetric stretch and symmetric bend,  respectively,
whilst $v_3$ and $v_4$ are the asymmetric stretch and asymmetric bend
respectively. The additional quantum numbers $l_3$ and $l_4$ are necessary
to resolve the degeneracy of the $v_3$ and $v_4$ vibrational states,
respectively.

The selection rules which determine the allowed electric dipole transitions
of $^{14}$NH\3\ are $\Delta J = J\p-J\pp = 0, \pm 1$ ($J\pp+J\p \ge 1$) with
symmetry selection rules, $A_2\p \leftrightarrow A_2\pp$, and $ E\p
\leftrightarrow E\pp$. We used the nuclear spin statistical weight factor
$g_{\rm ns}$  = 12 and 6 for the $A_2\p \leftrightarrow A_2\pp$ and $ E\p
\leftrightarrow E\pp$ transitions, respectively. The $A_1\p$ and $A_1\pp$
levels are characterized by $g_{\rm ns} = 0$, that is, the corresponding
transitions do not exist.

It should be noted that our assignments do not always agree with the
experimental ones for the following reasons: (i) ambiguous definition
of the quantum numbers (apart from $J$ and $\Gamma$), which depend on
the basis functions used; (ii) strong interactions between
ro-vibrational states of close-lying levels; and (iii) mapping between
the normal and local mode labels is not always straightforward. The
last of these means that it is sometimes difficult to distinguish
between the symmetric and asymmetric stretch quantum numbers $v_1$ and
$v_3$.

\begin{table}
\caption{\label{t:Energy-file} Extract from the BYTe Energy file.}
\scriptsize \tabcolsep=5pt
\renewcommand{\arraystretch}{1.0}
$
\begin{array}{rrrrrrrrrrrrrrrrrrrrrr}
    \hline
    \hline
1    &  2    &  3        &  4    &         5       &    6  &  7   &  8  &  9  &  10   &  11   &  12 &  13  &  14  &  15  &  16  &  17  &  18  &  19  &  20  &  21  &  22 \\
N    &  J    &  \Gamma   &  N_{\rm block} &  {\rm Term value} &
                                                      n_1  &  n_2  &  n_3  &    n_4&  l_3  & l_4  &  \tau_{\rm inv}
                                                                                                           &     J &     K &  \tau_{\rm rot} &  v_1  &  v_2  &  v_3  &  v_4  &  v_5  &  v_6  &  \Gamma_{\rm vib} \\
\hline
   1 &     0 &         1 &     1 &        0.000000 &     0 &     0 &     0 &     0 &     0 &     0 &     0 &     0 &     0 &     0 &     0 &     0 &     0 &     0 &     0 &     0 &     1 \\
   2 &     0 &         1 &     2 &      932.438362 &     0 &     1 &     0 &     0 &     0 &     0 &     0 &     0 &     0 &     0 &     0 &     0 &     0 &     0 &     0 &     2 &     1 \\
   3 &     0 &         1 &     3 &     1597.487235 &     0 &     2 &     0 &     0 &     0 &     0 &     0 &     0 &     0 &     0 &     0 &     0 &     0 &     0 &     0 &     4 &     1 \\
   4 &     0 &         1 &     4 &     2384.162466 &     0 &     3 &     0 &     0 &     0 &     0 &     0 &     0 &     0 &     0 &     0 &     0 &     0 &     0 &     0 &     6 &     1 \\
   5 &     0 &         1 &     5 &     3215.999346 &     0 &     0 &     0 &     2 &     0 &     0 &     0 &     0 &     0 &     0 &     0 &     0 &     0 &     0 &     2 &     0 &     1 \\
   6 &     0 &         1 &     6 &     3336.068662 &     1 &     0 &     0 &     0 &     0 &     0 &     0 &     0 &     0 &     0 &     0 &     1 &     0 &     0 &     0 &     0 &     1 \\
   7 &     0 &         1 &     7 &     3462.472133 &     0 &     4 &     0 &     0 &     0 &     0 &     0 &     0 &     0 &     0 &     0 &     0 &     0 &     0 &     0 &     8 &     1 \\
   8 &     0 &         1 &     8 &     4115.619903 &     0 &     1 &     0 &     2 &     0 &     0 &     0 &     0 &     0 &     0 &     0 &     0 &     0 &     0 &     2 &     2 &     1 \\
   9 &     0 &         1 &     9 &     4294.521388 &     1 &     1 &     0 &     0 &     0 &     0 &     0 &     0 &     0 &     0 &     0 &     1 &     0 &     0 &     0 &     2 &     1 \\
  10 &     0 &         1 &    10 &     4695.218516 &     0 &     5 &     0 &     0 &     0 &     0 &     0 &     0 &     0 &     0 &     0 &     0 &     0 &     0 &     0 &    10 &     1 \\
\hline
\hline
\end{array}
 $
\begin{tabular}{cll}
             Column           &    Notation                 &      \\
\hline
  $            1             $&$          N                $&    Level number (row)    \\
  $            2             $&$          J                $&    Rotational quantum number, angular momentum  \\
  $            3             $&$   \Gamma                  $&    Total symmetry in D$_{3h}$(M)                 \\
  $            4             $&$   N_{\rm block}           $&    Level number in a  block                      \\
  $            5             $&$   E                       $&    Term value (in \cm)                           \\
  $        6 \ldots 9        $&$   n_1 \ldots n_4          $&    Normal mode vibrational quantum numbers       \\
  $           10,11          $&$   l_3, l_4                $&    Vibratiobal angular momenta,  normal mode vibrational quantum \\
  $             12           $&$   \tau_{\rm inv}          $&    Inversional parity (0,1)                                                    \\
  $             13           $&$           J               $&    Rotational quantum number (the same as column 2)                             \\
  $             14           $&$           K               $&    Rotational quantum number, projection of $J$ onto the $z$-axis                \\
  $             15           $&$   \tau_{\rm rot}          $&    Rotational parity (0,1)                                                        \\
  $        16\ldots 21       $&$   v_1 \ldots v_6          $&    Local mode vibrational quantum numbers (see \protect\citep{NH3-T300K-paper}) \\
  $             22           $&$   \Gamma_{\rm vib}        $&    Symmetry of the vibrational contribution in D$_{3h}$(M) \\
\hline
\end{tabular}
\end{table}

\begin{table}
\caption{\label{t:Transit-file} Extract from the BYTe Transition file.}
\begin{center}
\scriptsize \tabcolsep=5pt
\renewcommand{\arraystretch}{1.0}
\begin{tabular}{rrr}
    \hline
    \hline
        $N\pp $      &   $N\p  $         &  A$_{\rm if}$ / s$^{-1}$ \\
    \hline
        \verb!8851!  &      \verb!5949!  &  \verb!1.56092538E-03! \\
        \verb!8852!  &      \verb!5949!  &  \verb!5.48772171E+00! \\
        \verb!8853!  &      \verb!5949!  &  \verb!3.32423612E-01! \\
        \verb!8854!  &      \verb!5949!  &  \verb!2.71613285E+00! \\
        \verb!8855!  &      \verb!5949!  &  \verb!4.45088305E-02! \\
    \hline
\end{tabular}
\end{center}
\end{table}


\begin{figure}
\begin{center}
\includegraphics[width=0.6\linewidth]{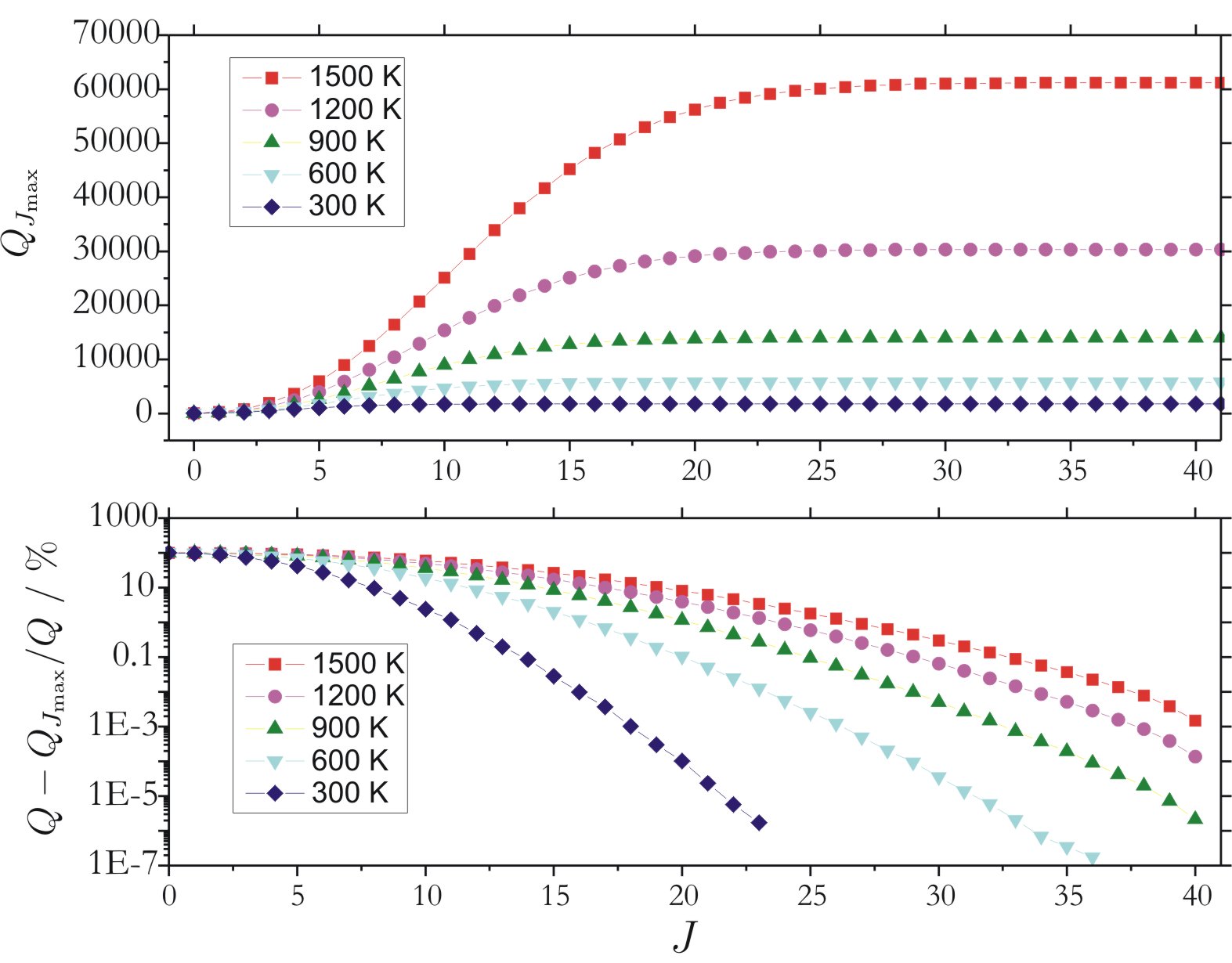}
\end{center}
\caption{\label{f:partfunc} The partition functions $Q_{J_{\rm max}}$ of NH\3\ at different temperatures \textit{vs.}
the maximum $J$ value used in eq.~(\ref{eq:pf}), $J_{\rm max}$, (upper display)
and corresponding contributions $(Q - Q_{J_{\rm max}})/Q$ (\%), where $Q = Q_{ J_{\rm max} = 41   }$.  }
\end{figure}

\begin{figure}
\begin{center}
\includegraphics[width=0.6\linewidth]{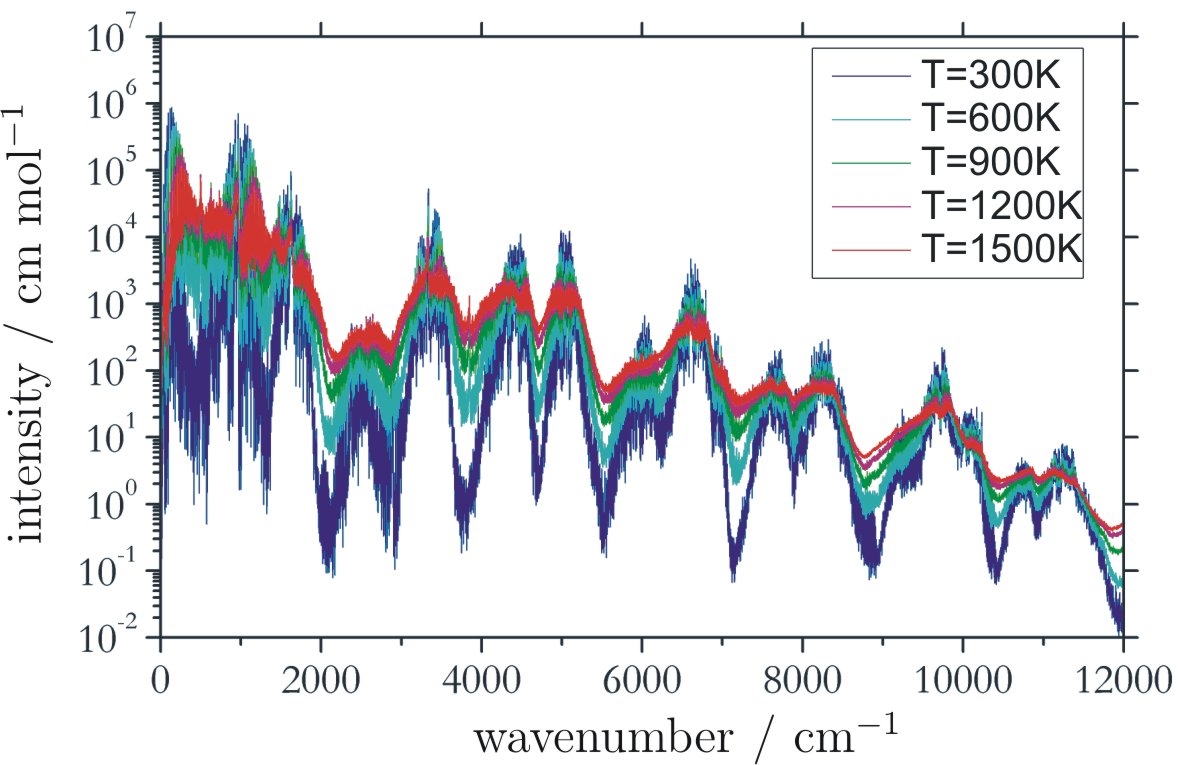}
\end{center}
\caption{\label{f:absorp:total:diff-T} Absorption spectra of NH\3\ given by BYTe for
$T=300$, 600, 900, 1200, and 1500~K, convoluted with a Gaussian
profile, HWHM = 0.5 cm$^{-1}$. }
\end{figure}

\begin{figure}
\begin{center}
\includegraphics[width=0.8\linewidth]{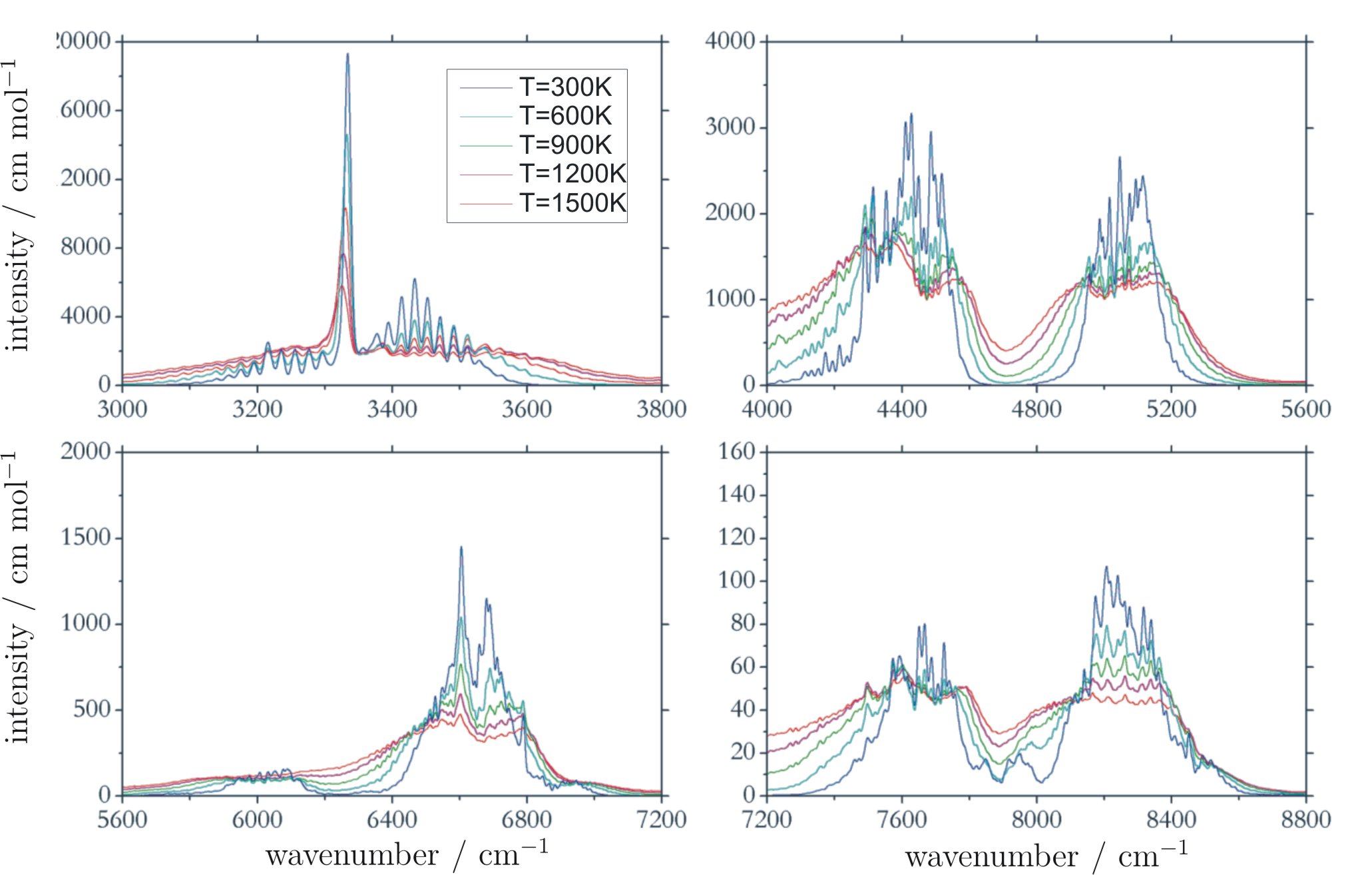}
\end{center}
\caption{\label{f:absorp:ABCD}BYTe absorption spectra of NH\3\ for $T=300$,
600, 900, 1200, and 1500~K, convoluted with a Gaussian profile, HWHM =
5 cm$^{-1}$: The 3~$\mu$m, 2~$\mu$m, 1.5~$\mu$m and 1.25~$\mu$m
regions. }

\end{figure}

\begin{figure}
\begin{center}
\includegraphics[width=0.6\linewidth]{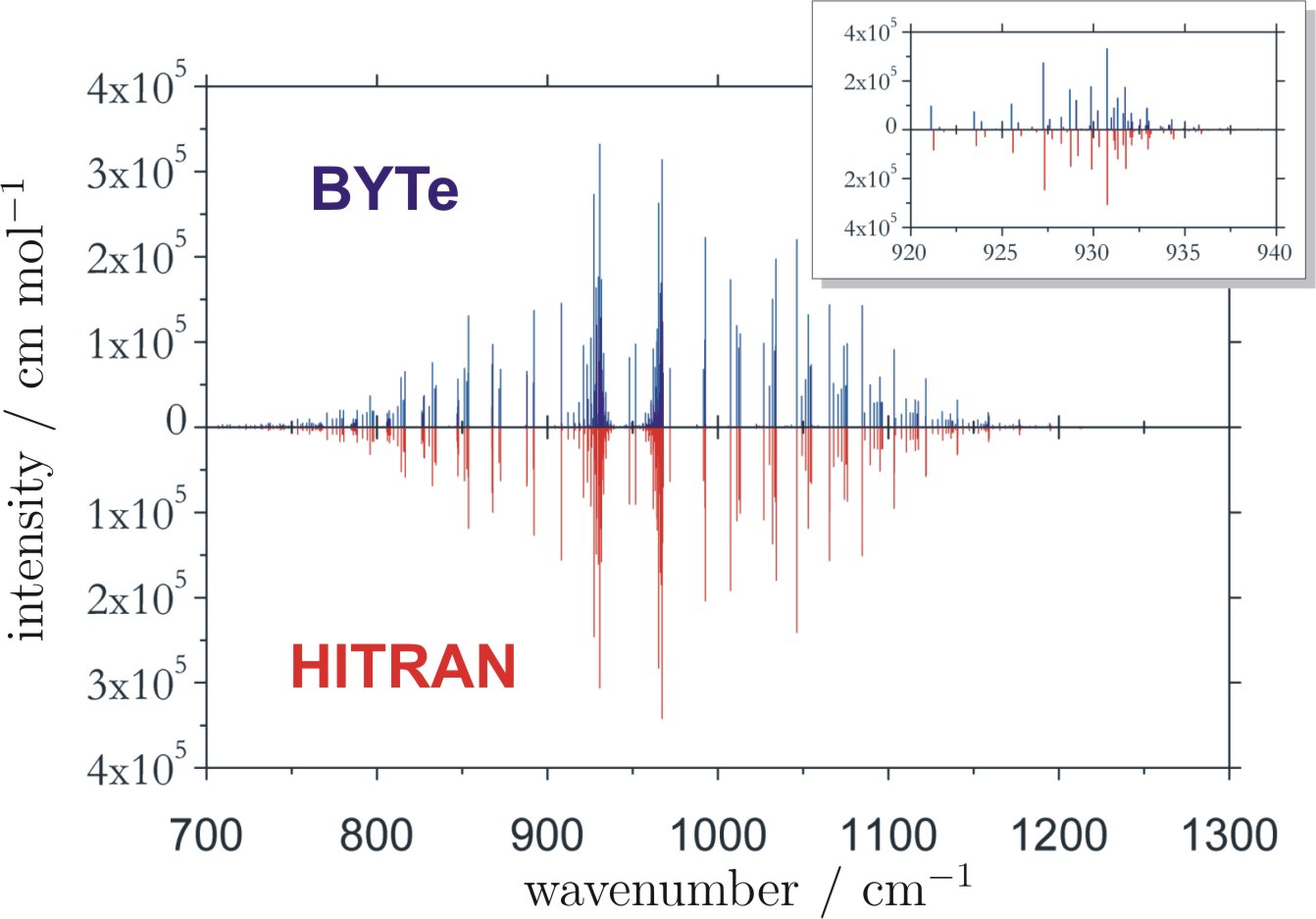}
\end{center}
\caption{\label{f:absorb:T300K} Absorption spectra of NH\3\ at T=300~K  (cm/mol): BYTe \textit{vs} HITRAN.}
\end{figure}

\begin{figure}
\begin{center}
\includegraphics[width=0.6\linewidth]{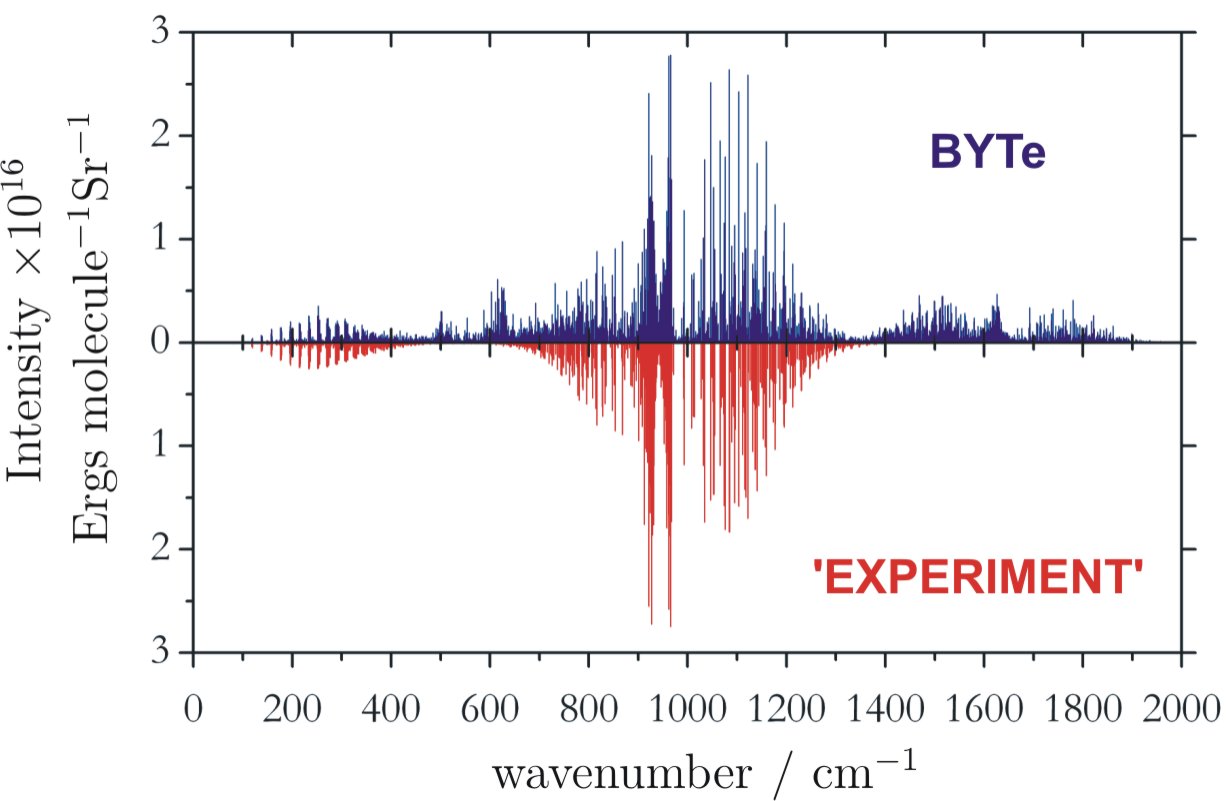}
\end{center}
\caption{\label{f:emiss:T900K} Emission spectra of NH\3\ at T=900~K  (Ergs/molecule/Sr). The `experimental spectrum' (lower display)
is reconstructed using the Einstein coefficients from \protect\citet{Yu-NH3-T900K}.}
\end{figure}

\section{Line list: Description and Validation}
\label{s:ll-validation}

Absorption and emission intensity simulations are
temperature-dependent. Specifically, temperature appears in the
Boltzmann factors $\exp(-E/ kT)$, where $k$ is the Boltzmann
constant. Line lists, in contrast, do not specify a temperature, since
the Einstein coefficients for the transitions are independent of
temperature. We therefore need to explain why we refer to BYTe as a
`hot' list. The reason is that BYTe is able to describe the
absorption/emission processes in NH\3\ for temperatures up to $T_{\rm
max} = 1\,500$~K. This is because the energy threshold that we have
adopted, $E_{\rm low} = $ 8\,000~\cm, ensures that we include all
those states that are significantly populated up to  $T=T_{\rm
max}$.

The temperature coverage of our data can be conveniently checked by
using it to compute temperature-dependent partition functions for NH$_3$.

\begin{equation}
Q = \sum_j g_j \, \exp (  -E_j/kT),
\label{eq:pf}\end{equation}
where $g_j$ is the total degeneracy of the state with energy $E_j$ and
the sum runs over all energy levels of the molecule. Including all
energy levels up to 8\,000~\cm\ gives an NH\3\ partition function,
$Q$, of 57\,944 at $T=$1\,500~K. This value is about 5~\% below our
best estimate for $Q$ at 1\,500 of 61\,223. This discrepancy, which
rapidly disappears at lower temperatures, is the reason we do not
recommend the use of BYTe for temperatures above 1\,500~K.

Apart from $E_{\rm low}$ there are two main contributions that affect
the total number of the excited ro-vibrational states accessed in the
variational calculations, (i) the size of the vibrational basis set,
which BYTe is determined by the polyad number $P_{\rm max}$ (see
Eq.\ref{e:polyad-2}) and (ii) the size of the rotational basis set,
determined by the maximal total angular momentum $J_{\rm
max}$. Fig.~\ref{f:partfunc} illustrates how the partition function of
NH$_3$ depends on $J_{\rm max}$ for $T =$ 300~K, 600~K, 900~K, 1200~K,
and 1500~K (upper display). For $T=1\,500$~K the contribution to $Q$
from $J > 30$ amounts to less than 1~\%, see lower display of
Fig.~\ref{f:partfunc}, where the relative contributions $(Q -
Q_{J_{\rm max}})/Q$ to the partition function are also shown as
percentages. Here we used $Q = Q_{J_{\rm max} = 41}$, where $J_{\rm
max}$ = 41, corresponding to the highest $J$ in our variational
calculations.

The effect of the vibrational basis set on the partition function is
less pronounced: at $T=1\,500$~K and $J_{\rm max} = 35$ the partition
function is converged to better than 0.03~\% when testing our choice
of basis set size.   There are, of
course, other factors that give rise to errors in our computed partition
functions. One of these is the potential energy function, which is only
reasonably accurately fitted to low-lying ro-vibrational levels. We
will address the nature of the partition function of ammonia in
greater details elsewhere.

Figs.~\ref{f:absorp:total:diff-T} and \ref{f:absorp:ABCD} illustrate
the $T$-dependence of the absorption spectra of ammonia, and give an
overview of the complete range as well of the four selected regions
computed at $T =$ 300~K, 600~K, 900~K, 1\,200~K, and 1\,500~K. As
expected, the spectrum profiles at higher $T$ become less extreme as
the populations from the lower vibrational states are reduced in
favour of the vibrationally excited states. Only those few bands that
are the most pronounced features at $T=300$~K are still recognizable
at $T=1\,500$~K.

Our `cold' NH$_3$ paper \citep{NH3-T300K-paper} contained detailed comparisons
with the HITRAN database \citep{HITRAN} which demonstrated the accuracy
of our procedure and also, that even at 300~K, HITRAN is missing
significant NH\3\ data. In BYTe we have improved our ability to
reproduce the HITRAN data, see Fig.~\ref{f:absorb:T300K}; however we
do not show the entire comparison here.

As we were completing our calculations, a high temperature emission
spectra of ammonia ($T=900$~K) for the ground and $\nu_2$ states of
ammonia was reported by \citet{Yu-NH3-T900K}.  We use this spectrum to
provide an independent validation of the BYTe line list.  To this end
we used BYTe to generate a 900 K synthetic emission spectrum of NH\3\
in the same spectral region. This is shown in Fig.~\ref{f:emiss:T900K}
(lower part). The `experimental' spectrum on this figure was generated
using the Einstein coefficients from the synthetic line list reported
by \citet{Yu-NH3-T900K}.  The spectra agree not only qualitatively but
also quantitatively in terms of the absolute intensity values. The
additional lines in the upper part of Fig.~\ref{f:emiss:T900K} are
present because our synthetic spectrum includes all possible
transitions falling into the region including hot bands, while Yu {\it
et al}'s data is only for the
ground and $\nu_2$ states of NH\3.

\section{Conclusion}
\label{s:conclusion}

We have calculated to a high level of accuracy the frequencies and
Einstein $A$ coefficients of all the transitions that are present in
the emission and absorption spectrum of NH\3. The only limitations
are: upper states with energies above 18\,000~\cm\ are excluded;
there is an effective short-end wavelength cut-off of 1 $\mu$m, due to
the incompleteness of our data at frequencies above 10\,000 \cm\ (the
excluded region is unimportant for NH\3), and extremely weak lines
have been excluded, which is of little physical significance.

Although the BYTe line list is explicitly aimed at modelling hot ammonia,
it improves on our previous cold line list \citep{NH3-T300K-paper}
in terms of the quality of the potential energy surface used, in the
size of the basis sets employed and in the range of frequencies studied.
We therefore recommend the use of BYTe for all temperatures up to
1\,500~K. The line list is freely available and can download in its entirety
or in parts from:
{\it http://www.tampa.phys.ucl.ac.uk/ftp/astrodata/BYTe}.

\section*{Acknowledgments}
 We thank members of UCL's Research Computing who have supported this
work by giving us extended access to both the Legion and Unity high
performance computing systems. We are also grateful to Shanshan Yu for
providing her results prior to publication. We thank the Leverhulme
Trust for funding this work and STFC for support via a grant to the
"Miracle" computing consortium.

\bibliographystyle{mn2e}

\label{lastpage}

\end{document}